# Mapping active allosteric loci SARS-CoV Spike Proteins by means of Protein Contact Networks

*Luisa Di Paola[1] and Alessandro Giuliani[2]*

1. Unit of Chemical-physics Fundamentals in Chemical Engineering, Department of Engineering, Università Campus Bio-Medico di Roma, via Álvaro del Portillo 21, 00128 Rome, Italy
2. Environmental and Health Department, Istituto Superiore di Sanità, 00161 Rome, Italy

**ABSTRACT**

Coronaviruses are a class of virus responsible of the recent outbreak of Human Severe Acute Respiratory Syndrome. The molecular machinery behind the viral entry and thus infectivity is based on the formation of the complex of virus spike protein with the angiotensin-converting enzyme 2 (ACE2). The detection of putative allosteric sites on the viral spike protein can trace the path to develop allosteric drugs to weaken the strength of the spike-ACE2 interface and, thus, reduce the viral infectivity.

In this work we present results of the application of the Protein Contact Network (PCN) paradigm to the complex SARS-CoV spike – ACE2 relative to both 2003 SARS and the recent 2019 – CoV. Results point to a specific region, present in both structures, that is predicted to act as allosteric site modulating the binding of the spike protein with ACE2.

**Keywords**: SARS-CoV spike protein, ACE2 binding, Protein Contact Networks , allosteric drug.

**INTRODUCTION**

Coronaviruses are large, enveloped, positive stranded RNA viruses, causing gastrointestinal, nervous system and respiratory distresses [1]. The novel coronavirus emerged in 2019 in the region of Wuhan, China, shows many similarities with other forms, but also some specific features [2]. The formulation of effective therapies and vaccines requires the knowledge of the molecular mechanisms underlying virus infection and diffusion.

The novel coronavirus 2019-nCoV belongs to the *Orthocoronavirinae* subfamily, the same of the MERS-CoV and SARS-CoV. It has been recognized as distinct member of the subfamily in the late 2019 [3] and its spreading in China and in several other countries it's a major concern for the global health.

The molecular mechanism of the viral infection of all coronaviruses involves the spike protein S, a trimeric, petal-like protein protruding from the viral capsid. It interacts with a host cell receptor, the angiotensin converting enzyme 2 (ACE2) to ignite viral fusion with the host cells, the very first stage of the viral infection [4,5].

Understanding the formation of the complex of spike protein with its receptor, the ACE2, may trace the path to develop therapies and vaccines for the coronaviruses [6], based on the inhibition of the formation of binding interface between viral S and host ACE2 proteins.

In this perspective, an allosteric modulation strategy is more promising than an orthosteric one. Drugs interacting with allosteric sites (allosteric modulators [7]) target more accessible protein regions with respect to the orthosteric drugs directly interfering with interaction (or in general active) sites. Actually, active (interaction) sites, in order to preserve the specificity of the reaction (interaction), are often tightly packed and sterically hindered. In this sense, the development of allosteric drugs is a hot topic in the drug discovery field, allowing also to rethink the druggability of undruggable proteins [8].

In this work, we try to identify the most promising 'druggable' allosteric site of viral protein S by means of the application of the Protein Contact Network (PCN), an emerging paradigm in computational biochemistry [9–11], to two complexes of spike proteins with the ACE2 receptor. The application of the method clearly highlights regions in the spike protein acting as putative *loci* for allosteric modulation.

**MATERIALS AND METHODS**

The analyzed structures are: complex SARS-spike glycoprotein – human ACE2 complex (stabilized variant, all ACE2-bound particles, PDB code 6CS2 [12], from now on SARS-CoV S/ACE2 complex) and the nCoV 2019 analogous spike glycoprotein- human ACE2 complex [13] (available on the site https://zhanglab.ccmb.med.umich.edu/C-I-TASSER/2019-nCov/ code S_ACE2.pdb, from now on nCoV2019 S/ACE2 complex ).

A purposed software transformed the full structural information in the PDB files into a Protein Contact Network (PCN): network nodes are the protein residues, represented by alpha-carbons, links between nodes (residues) exist if the residues mutual distance lies in the range between 4 and 8 Å, thus including only significant ( <8 Å) noncovalent (>4 Å) bonds.

The adjacency matrix *A* represents mathematically the PCN in terms of undirected, unweighted network and it is defined as:

$$A_{ij} = \begin{cases} 1 & if\ 4\ \text{Å} < d_{ij} < 8\ \text{Å} \\ 0 & otherwise \end{cases} \quad (1)$$

The topological role of nodes (residues) address functional role at the corresponding residues, on the basis of value of network descriptors. The method is widely discussed elsewhere [14].

The basic network descriptor is the node degree $k_i$, defined as:

$$k_i = \sum_j A_{ij} \quad (2)$$

To detect allosteric sites and, more in general, functional regions activating upon binding, we adopted a method based on network spectral clustering [15,16]. Spectral clustering is based on the Laplacian matrix spectral decomposition, where the Laplacian matrix $\mathbf{L}$ is defined as:

$$\mathbf{L} = \mathbf{D} - \mathbf{A} \qquad (3)$$

$\mathbf{D}$ is the degree matrix, i.e. a diagonal matrix whose diagonal is the degree vector.

The spectral clustering is based on the eigenvalue decomposition of the Laplacian $\mathbf{L}$. The sign of the Fiedler vector (the eigenvector corresponding to the second minor eigenvalue) is used for binary partition (each network is parted into two clusters, which in turn can be parted into four, and so on), as outlined in [16,17]. Network clusters (group of residues) correspond to functional regions in protein [16].

Once network nodes (again, protein residues) are parted into the given number of clusters (independently assigned at the beginning of the clustering procedure), it is possible to define the participation coefficient, defined as:

$$P_i = 1 - \left(\frac{k_{si}}{k_i}\right)^2 \qquad (4)$$

where $k_{si}$ is the degree of the i-th residue computed in the cluster it belongs (i.e. accounting only for links with nodes pertaining to the same cluster).

This network descriptor has been demonstrated to represent the functional role of residues in binding and stability [14,16,18,19]. Residues with high value of $P$ are responsible for communication between clusters (functional regions) and thus are addressed with a role in allosteric communication [20]. In this work, the participation coefficient maps will be projected onto ribbon protein structures (as heat maps), to highlight hotspots activating in the spike-ACE2 complex. Participation coefficient maps are visualized by PyMol (https://pymol.org/2/).

We characterized the spike protein/ACE2 interface by means of descriptors mediated by the structural information and the network analysis:

1. the interchain degree $k_i^{IC}$ defined as the node degree, but only for nodes (residues) belonging to different chains;
1. the interface roughness $Q/R$ [21] defined for each chain participating in an interface, $Q$ is the number of chain residues in the interface and $R$ the sequence range of chain residues in the interface;
2. the interface amino acid range [21] $IAR=R/N$, $N$ is the total number of residues in the chain;
3. the interface energy matrix $E$ defined as:

$$E = E_{ij} = \begin{cases} e_{ij} = \frac{1}{d_{ij}} & \text{if } 4\,\text{Å} < d_{ij} < 8\,\text{Å and the residues belong to different chains} \\ 0 & \text{otherwise} \end{cases} \qquad (5)$$

the interface energy $E_{INT}$ is the sum of $e_{ij}$ and $\langle E_{INT} \rangle$ is the value averaged over the whole number of residues at the interface

We computed general interface properties of the spike/ACE2 complex by the PISA software [22], specifically the interface area $S_{INT}$ and the energy gain upon the interface formation $\Delta G_{INT}$.

**RESULTS AND DISCUSSION**

Figure 1 reports the cluster partition (2 clusters) SARS-CoV S/ ACE2 complex (Fig1B); the chains are also shown to highlight which chains participate in the two clusters (Fig 1A).

The interface between the fusion peptide and the ACE2 ectodomain is so strong (in terms of number of contacts between viral and host proteins) that the clustering partition algorithm recognized the fusion+ACE2 region as a single cluster, even though made up of sequences relative to different chains.

The map of the participation coefficient projected onto the ribbon structure of the SARS-CoV/ACE2 complex (Fig 1C) shows an active region (P > 0) in the junction between the fusion peptide and the trimeric bulk phase of the spike protein. Active residues are shown more in details in red in panel 1D: as previously demonstrated, the participation coefficient describes the attitude of residues to participate to intercommunication between clusters, so it is a putative descriptor of the allosteric communication of residues. This region is thus a good candidate to intervene in the allosteric regulation of the complex formation and, therefore, it is worth to investigate allosteric drugs targeting this region.

Figure 2 refers to the analogous analysis for the nCov2019 S/ ACE2 complex.

A similar active region appears at the junction of the fusion peptide to the body of the spike protein, yet the active region in the SARS CoV S/ACE2 is more compact than in the nCoV2019 S/ACE2. Active regions in both complexes are characterized by two beta-sheets and unfolded traits, thus liable to be targeted to peptides due to flexibility of the region and the absence of a pocket to bind small organic molecules [23].

Table 1 reports the value of the participation coefficient *P* in the residues with *P > 0*, all located in the chain carrying the fusion peptides of SARS – CoV S/ACE2 (chain B, in cyan in Figure 1.A) and nCoV2019 S/ACE2 (chain C, in magenta in Figure 1.A) complexes.

The SARS CoV S/ACE2 complex accounts for 21, the nCoV2019 S/ACE2 for 22 active residues. The average value is similar for the two distributions (0.47 for SARS CoV S/ACE2, 0.48 for nCoV2019 S/ACE2). However, the absolute values of higher P residues ( >0.6 ) are higher for the nCoV2019 S/ACE2, pointing to a corresponding larger reactivity of the active region.

Table 2 reports the properties of the interface spike/ACE2 in the two complexes.

The interface area and the absolute value of energy of the SARS CoV S/ACE2 complex are higher than the corresponding for the nCoV2019 S/ACE2 complex; however, the specific values for unit area and single residues are higher for nCoV2019 S/ACE2, pointing to a more efficient yet less stable interface in the complex.

**CONCLUSIONS**

The application of the method of Protein Contact Networks to two complexes of the spike protein S with its receptor, the angiotensin-converting enzyme ACE2, has disclosed active zones in the complex of SARS-CoV and 2019-nCoV with the ACE2 ectodomain.

 These regions are worth of further investigation in the perspective of drug and vaccine development since, as demonstrated in previous works [14,24,25], they have a very high probability to be the allosteric sites in charge of the modulation the spike/ACE2 proteins binding.

**Figure 1** - SARS-CoV S/ACE2 complex (PDB code 6cs2 [12]). A) partition into two clusters (green cluster 1, comprising the ectodomain of ACE2 and the fusion peptide, in red cluster 2, the remaining part of the spike protein); B) chains in the complex, in green the chain A, in cyan the chain B, carrying the fusion peptide, in magenta the chain C, in yellow the ACE2 ectodomain.

**Figure 2** – nCoV2019 S/ACE2 complex (PDB code 6cs2 [1]). A) partition into two clusters (green cluster 1, comprising the ectodomain of ACE2 and the fusion peptide, in red cluster 2, the remaining part of the spike protein); B) chains in the complex, in green the chain A, in cyan the chain B, in magenta the chain C, carrying the fusion peptide, in yellow the ACE2 ectodomain.

**Table 1** - Values of the participation coefficient P for the SARS-CoV S/ACE2 and nCoV2019 S/ACE2 complexes (only positive values are reported, with corresponding position, in italic values larger than 0.6).

**Table 2** - Properties of the spike protein / ACE2 interface.

FIGURE 1

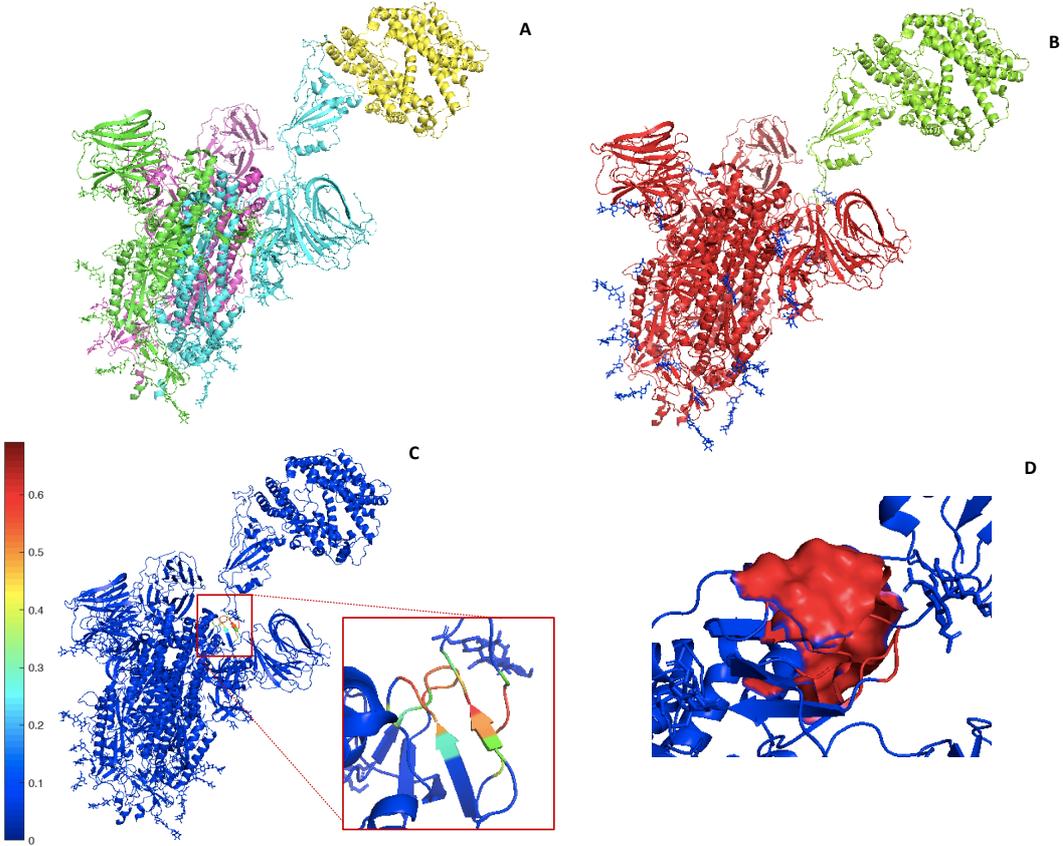

FIGURE 2

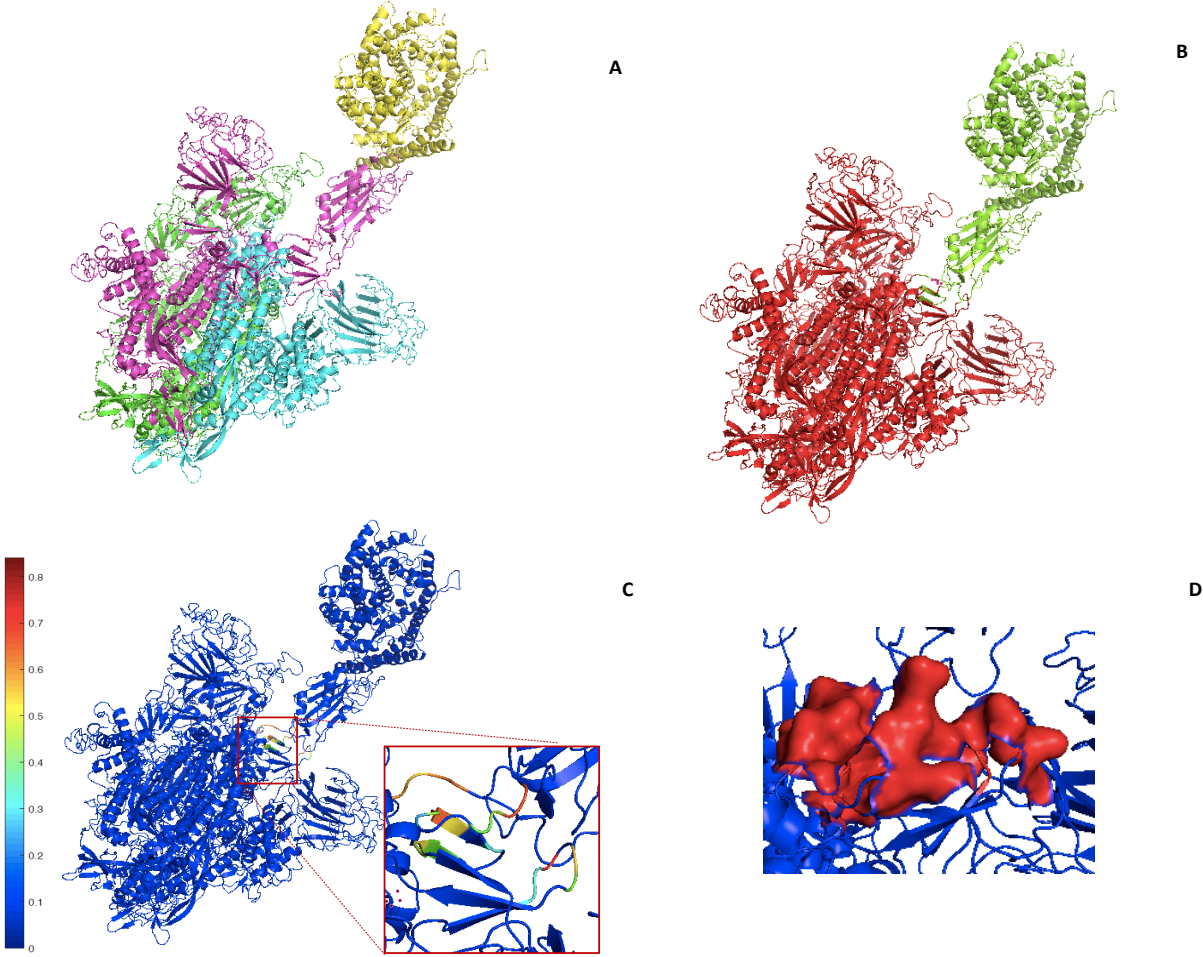

TABLE 1

| SARS CoV S/ACE2 | | nCoV2019 S/ACE2 | |
|---|---|---|---|
| aa | P | aa | P |
| 311D | 0.31 | *320Q* | *0.61* |
| 312V | 0.56 | 321P | 0.56 |
| *313V* | *0.69* | *322T* | *0.64* |
| 314R | 0.47 | *323E* | *0.67* |
| 315F | 0.23 | *531N* | *0.75* |
| 515S | 0.31 | *532L* | *0.75* |
| *516T* | *0.64* | 533V | 0.44 |
| *517D* | *0.64* | 534K | 0.31 |
| 518L | 0.56 | 537C | 0.15 |
| 519I | 0.36 | 538V | 0.44 |
| 526F | 0.19 | *539N* | *0.70* |
| 527N | 0.51 | 540F | 0.51 |
| *528N* | *0.61* | 542F | 0.21 |
| *529N* | *0.64* | 546T | 0.31 |
| 531L | 0.36 | 547G | 0.36 |
| 562R | 0.21 | 548T | 0.51 |
| 563D | 0.27 | 549G | 0.44 |
| *564P* | *0.61* | 576R | 0.21 |
| 565K | 0.56 | 577D | 0.21 |
| 566T | 0.56 | *578P* | *0.84* |
| *567S* | *0.64* | 580T | 0.56 |
| - | - | 581L | 0.36 |

TABLE 2

|  | **SARS-CoV SA/ACE2** | **nCoV2019 S/ACE2** |
|---|---|---|
| $IAR_{SP}$ | 0.01 | 0.008 |
| $IAR_{ACE}$ | 0.62 | 0.56 |
| $(Q/R)_{SP}$ | 0.45 | 0.37 |
| $(Q/R)_{ACE}$ | 0.05 | 0.04 |
| $E_{INT}$ | 5.37 | 4.35 |
| $<E_{INT}>$ | 0.17 | 0.17 |
| $n_{IC}=Q_{SP}+Q_{ACE}$ | 31 | 26 |
| $S_{INT}$, Å$^2$ | 962.2 | 739.9 |
| $\Delta G_{INT}$, kcal/mol | -7.9 | -7.4 |
| $(\Delta G_{INT}/S_{INT})$, cal/(mol Å$^2$) | 8.21 | 10.00 |